\documentclass[a4paper,12pt]{article}
\pdfoutput=1
\usepackage{amssymb}
\usepackage{epsfig}
\usepackage{graphicx}
\usepackage{slashed}

\usepackage[utf8]{inputenc}

\makeatletter

\@addtoreset{equation}{section}
\makeatother
\def\be{\begin{equation}}
\def\ee{\end{equation}}
\def\bea{\begin{eqnarray}}
\def\eea{\end{eqnarray}}

\def\({\left(}
\def\){\right)}
\def\<{\left<}
\def\>{\right>}

\def\be{\begin{equation}}
\def\ee{\end{equation}}
\def\bea{\begin{eqnarray*}}
\def\eea{\end{eqnarray*}}
\def\ben{\begin{eqnarray}}
\def\een{\end{eqnarray}}
\def\({\left(}
\def\){\right)}
\def\<{\left<}
\def\>{\right>}
\def\!{\right|}
\def\|{\left|}

\def\[{\left[}
\def\]{\right]}

\def\+{\bar}

\def\D{{\bf{D}}}
\def\L{{\cal{L}}}
\def\t{\widetilde}

\def\F{{\cal{F}}}

\def\P{{\cal{P}}}

\def\L{{\cal{L}}}

\def\eps{{\cal{\varepsilon}}}

\def\E{{\cal{E}}}

\def\F{{\cal{F}}}

\def\h{\widehat}

\begin{document}

\setlength{\unitlength}{1mm}

\pagestyle{empty}
\vskip-10pt
\vskip-10pt
\hfill 
\begin{center}
\vskip 3truecm
{\Large \bf
A proposal for the non-Abelian tensor multiplet}
\vskip 2truecm
{\large \bf
Andreas Gustavsson}
\vspace{1cm} 
\begin{center} 
Physics Department, University of Seoul, 13 Siripdae, Seoul 130-743 Korea
\end{center}
\vskip 0.7truecm
\begin{center}
(\tt agbrev@gmail.com)
\end{center}
\end{center}
\vskip 2truecm
{\abstract{If one compactifies the Abelian $(1,0)$ tensor multiplet on a circle, one finds 5d SYM for the zero modes. For the Kaluza-Klein modes one can likewise find a Lagrangian description in 5d \cite{Bonetti:2012st}. Since in 5d we have an ordinary YM gauge potential, one may look for a non-Abelian generalization and indeed such a non-Abelian generalization was found in \cite{Bonetti:2012st}. In this paper, we study this non-Abelian generalization for the $(1,0)$ tensor multiplet in detail. We obtain the supersymmetry variations that we close on-shell. This way we get the fermionic equation of motion and a modified selfduality constraint.}}

\vfill
\vskip4pt
\eject
\pagestyle{plain}

\section{Introduction}
To address the 6d non-Abelian tensor multiplet theory, either with $(1,0)$ supersymmetry or $(2,0)$ supersymmetry, one may consider the theory compactified on a circle where one may relate it with the dimensionally reduced 5d SYM theory. 

One proposal \cite{Douglas:2010iu}, \cite{Lambert:2010iw} has been that 5d SYM theory is exactly the same thing as 6d (2,0) theory compactified on a circle, the instanton particles in the 5d SYM could play the role of Kaluza-Klein modes. 

Another proposal \cite{Bonetti:2012st} has been to incorporate Kaluza-Klein modes in the 5d description of the non-Abelian tensor multiplet. 

Our original motivation to study the proposal of \cite{Bonetti:2012st} was to try to find an inconsistency as it is unsatisfactory to have two different proposals for the 6d tensor multiplet which appear to be in conflict with each other. We did not find any inconsistency, at least not so far. In this paper we clarify part of the result in \cite{Bonetti:2012st}. For the sake of simplicity, we restrict ourselves to the non-Abelian $(1,0)$ tensor multiplet. We make an ansatz based on \cite{Bonetti:2012st} for the on-shell supersymmetry variations. We close these supersymmetry variations on-shell up to a gauge transformation. By doing this, we find closure relations on the following form
\bea
\delta^2 \Phi^{(n)} &=& - i s^m \partial_m \Phi^{(n)} - \frac{n s}{r} \Phi^{(n)}\cr
&& + \delta_{gauge} \Phi^{(n)} \cr
&& + ({\mbox{equation of motion}})
\eea
for all the fields in the non-Abelian $(1,0)$ tensor multiplet. The Kaluza-Klein mode number is $n$. The 5d parameters that form a 5d vector $s^m$ together with the 5d parameter $s$ have an uplift to a 6d Lorentz vector $S^M$, and the first line in the closure relation can be written in the 6d Lorentz covariant form as 
\bea
- i s^m \partial_m \Phi^{(n)} - \frac{n s}{r} \Phi^{(n)} &=& - i S^M \partial_M \Phi^{(n)}
\eea
where, if we compactify on a temporal circle of radius $r$, we have $\partial_0 \Phi^{(n)} = i n \Phi^{(n)}$. This closure relation suggests that a hidden 6d Lorentz symmetry may emerge in the decompactification limit. The second thing we learn is what the fermionic equation of motion shall be and the third thing we learn is that we need to modify the selfduality constraint in nontrivial way in order to close the supersymmetry variations on-shell. 

In principle it should be possible to derive the fermionic equation of motion and the modified selfduality constraint from the Lagrangian in \cite{Bonetti:2012st}. To this end, we would need to eliminate the auxiliary field from that Lagrangian. The auxiliary field $Y^I$ enters their Lagrangian as $\L = h_{IJ} Y^I Y^J + f_{IJK} \(\phi^I Y^J Y^K + Y^I \bar\chi^J \chi^K\) + ...$ where $f^{IJK}$ denote Lie algebra structure constants associated with the gauge group using their notation. Here we are suppressing the KK-mode numbers. If we would display the mode number, we would also see how $Y^I$ mixes zero modes and KK modes. To eliminate $Y^I$ from this Lagrangian, we need to invert the matrix $h_{IJ} + f_{IJK} \phi^K$. This seems like a challenging problem and it was not carried out in full generality in \cite{Bonetti:2012st}. Only two special cases where studied there; dimensional reduction resulting in the usual 5d SYM, and Abelian gauge group that puts $f^{IJK}=0$.

\section{The non-Abelian tensor multiplet}
As we outlined in the introduction, we will consider the 6d $(1,0)$ tensor multiplet on a circle. The classical 6d tensor multiplet is best understood in Lorentzian signature. We will  restrict ourselves to flat Lorentzian spacetime with the metric
\bea
ds^2 &=& - r^2 dt^2 + \delta_{mn} dx^m dx^n
\eea
We choose to compactify the time direction 
\bea
t &\sim & t + 2\pi
\eea
and reduce to Euclidean 5d SYM with the YM coupling constant 
\bea
g^2 &=& 4\pi^2 r
\eea
The 6d formulation of the non-Abelian tensor multiplet may not be known, but for the Abelian tensor multiplet we do know what it is. So we may start there. It consists of one scalar field $\sigma$, one fermionic field $\chi$ and a tensor field $B_{MN}$ with selfdual field strength. When we put this theory on a circle, we may expand the fields in modes as follows,
\bea
\sigma &=& \phi + \sum_{n\neq 0} \sigma^{(n)}\cr
\chi &=& \psi + \sum_{n \neq 0} \chi^{(n)}
\eea
where $n$ is integer, and 
\bea
\partial_t \sigma^{(n)} &=& i n \sigma^{(n)}
\eea
and similarly for the other fields. For the gauge field, we expand 
\bea
B_{m0} &=& a_m + \sum_{n\neq 0} A_m^{(n)}\cr
B_{mn} &=& b_{mn} + \sum_{n \neq 0} B_{mn}^{(n)}
\eea
To get the standard SYM normalization, we shall relate the zero modes to the SYM fields as $\phi_{SYM} = \phi/(2\pi r)$, $\psi_{SYM} = \psi/(2\pi r)$ and $(a_m)_{SYM} = a_m/(2\pi)$ but we will not make such a rescaling here. 

For the Abelian theory we define
\bea
\F_{mn}^{(n)} &=& \partial_m A_n^{(n)} - \partial_n A_m^{(n)} + i n B_{mn}^{(n)}\cr
H_{mnp}^{(n)} &=& \partial_m B_{np}^{(n)} + \partial_p B_{mn}^{(n)} + \partial_n B_{pm}^{(n)}
\eea
We have the selfduality constraints
\bea
H_{mnp}^{(n)} &=& -\frac{1}{2r}\E_{mnp}{}^{rs} \F_{rs}^{(n)}
\eea
and the Bianchi identities
\bea
3\partial_{[m} \F_{np]}^{(n)} - i n H_{mnp}^{(n)} &=& 0
\eea
We can use these conditions to eliminate $H_{mnp}^{(n)}$ and then we just need to work with $\F_{mn}$ subject to a `selfdual Bianchi identity'
\bea
3\partial_{[m} \F_{np]}^{(n)} + \frac{i n}{2r} \E_{mnp}{}^{rs} \F_{rs}^{(n)} &=& 0
\eea
It is now straightforward to derive the supersymmetry variations for these modes from the 6d supersymmetry variations. To this end, we need to fix some spinor conventions. We collect all our spinor conventions in Appendix $A$. For the zero modes, one finds
\bea
\delta a_m &=& i r \bar\eps\gamma_m \psi\cr
\delta \phi &=& - i \bar\eps \psi\cr
\delta \psi &=& \frac{1}{2r} \gamma^{mn} \eps f_{mn} + \gamma^m \eps \partial_m \sigma
\eea
and for the non-zero KK modes one finds
\bea
\delta \F_{mn} &=& - 2 i r \bar\eps\gamma_m D_n \chi - n \bar\eps \gamma_{mn} \chi\cr
\delta \sigma &=& - i \bar\eps \chi\cr
\delta \chi &=& \frac{1}{2r} \gamma^{mn} \eps \F_{mn} + \gamma^m \eps \partial_m \sigma + \frac{i n}{r} \eps \sigma 
\eea
We have suppressed the mode number $(n)$ which is common for all the KK fields that appear in the supersymmetry variations ($\F_{mn} = \F_{mn}^{(n)}$, $\chi = \chi^{(n)}$, $\sigma = \sigma^{(n)}$). This is nothing but a 5d reformulation of the Abelian 6d tensor multiplet, as we show explicitly in Appendix $B$.

Now once having this reformulation, it is natural to try to find a non-Abelian generalization by promoting the 5d gauge field $a_m$ to a non-Abelian gauge field and letting the other fields and the KK modes transform in the adjoint representation of the gauge group. For the zero mode part, we have the supersymmetry variations of 5d SYM
\ben
\delta a_m &=& i r \bar\eps\gamma_m \psi\cr
\delta \phi &=& - i \bar\eps \psi\cr
\delta \psi &=& \frac{1}{2r} \gamma^{mn} \eps f_{mn} + \gamma^m \eps D_m \sigma\label{SYM}
\een
For the KK-modes we make the following ansatz \cite{Bonetti:2012st}  
\ben
\delta \F_{mn} &=& - 2 i r \bar\eps\gamma_m D_n \chi - n \bar\eps \gamma_{mn} \chi\cr
&& + i C_1 [\phi,\bar\eps\gamma_{mn}\chi] - i C_1 [\sigma,\bar\eps\gamma_{mn}\psi]\cr
\delta \sigma &=& - i \bar\eps \chi\cr
\delta \chi &=& \frac{1}{2r} \gamma^{mn} \eps \F_{mn} + \gamma^m \eps D_m \sigma + \frac{i n}{r} \eps \sigma + \frac{C}{r} \eps [\phi,\sigma]\label{KK}
\een
for some parameters $C_1$ and $C$ that may depend on the mode number $n$. We define the non-Abelian YM field strength as
\bea
f_{mn} &=& \partial_m a_n - \partial_n a_m - i [a_m,a_n]
\eea
and the covariant derivative as
\bea
D_m \sigma &=& \partial_m \sigma - i [a_m,\sigma]
\eea
We now turn to the closure computation for these on-shell supersymmetry variations. Since the gauge parameter should agree for zero modes and nonzero modes, we will  obtain the closure relations for the zero mode fields first. 
 
\subsection{Closure on the SYM fields}
For the bosonic SYM fields, we get
\bea
\delta^2 \phi &=& - i s^m \partial_m \phi - i [\phi,\lambda_T]\cr
\delta^2 A_m &=& - i s^n \partial_n A_m + D_m \lambda_T
\eea
where the gauge parameter is 
\bea
\lambda_T &=& i s (a - r \phi)
\eea
Here we have introduced the vertical component of the gauge potential,
\bea
a &:=& a_m \frac{s^m}{s}
\eea
If we keep the covariant derivative, the closure relations take the form
\bea
\delta^2 \phi &=& - i s^m D_m \phi - i [\phi,\lambda]\cr
\eea
with 
\ben
\lambda &=& - i s r \phi\label{lambda}
\een
We have now determined the gauge parameter that should also appear in the closure for the KK modes, to which we now turn.

\subsection{Closure on $\sigma$}
\bea
\delta^2 \sigma &=& - i s^m D_m \sigma - \frac{s}{r} n \sigma - i [\sigma,\lambda]
\eea
where
\bea
\lambda &=& \frac{s C}{r} \phi
\eea
Matching this with (\ref{lambda}) determines 
\bea
C &=& - i r^2
\eea

\subsection{Closure on $\F_{mn}$}
\bea
\delta^2 \F_{mn} &=& - 2 i s^q D_n \F_{mq} - s^r \E_{mnr}{}^{pq} \(\frac{n}{2r}\F_{pq} - \frac{i C_1}{2 r} \([\sigma,f_{pq}] - [\phi,\F_{pq}]\)\) - \frac{ns}{r} \F_{mn}\cr
&& + \(\frac{iC_1}{r} - r\) s [f_{mn},\sigma]\cr
&& + 2 i \(C_1-C\) s_m D_n \([\phi,\sigma]\)\cr
&& - i \[\F_{mn},\frac{C_1 s}{r}\phi\]\cr
&& + {\mbox{fermionic bilinears}}
\eea
By taking 
\bea
C_1 = C = - i r^2
\eea
we get
\bea
\delta^2 \F_{mn} &=& - 2 i s^q D_n \F_{mq} - s^r \E_{mnr}{}^{pq} \(\frac{n}{2r} \F_{pq} - \frac{i C_1}{2 r} \([\sigma,f_{pq}] - [\phi,\F_{pq}]\)\) - \frac{ns}{r} \F_{mn}\cr
&& - i \[\F_{mn},\frac{C_1 s}{r}\phi\] + {\mbox{fermionic bilinears}}
\eea
We now read off the bosonic part of the selfdual Bianchi identity
\bea
3 D_{[r} \F_{mn]} + \frac{i n}{2 r} \E_{mnr}{}^{pq} \(\F_{pq} - \frac{r}{2} \([\sigma,f_{pq}]-[\phi,\F_{pq}]\)\) + {\mbox{fermions}}&=& 0
\eea
and the gauge parameter
\bea
\lambda &=& \frac{C_1 s}{r} \phi
\eea
Let us now look at the fermionic bilinears. These are
\bea
{\mbox{fermionic bilinears}} &=& - 2 i r^2 \{\bar\eps \gamma_n \psi,\bar\eps \gamma_m \chi\}\cr
&& + C_1 \{\bar\eps \psi,\bar\eps\gamma_{mn}\chi\}\cr
&& - C_1 \{\bar\eps \chi,\bar\eps\gamma_{mn}\psi\}
\eea
Now we use the identities
\bea
\gamma_n\gamma_p\gamma_m - \gamma_m \gamma_p \gamma_n &=& \{\gamma_{mn},\gamma_p\}\cr
\gamma_n \gamma_{pq} \gamma_m - \gamma_m \gamma_{pq} \gamma_n &=& - \{\gamma_{mn},\gamma_{pq}\}
\eea
We then use the following flipping rules,
\bea
\bar\eps\gamma_m\psi &=& \bar\psi\gamma_m\eps\cr
\bar\eps\psi &=& \bar\psi\eps
\eea
where we recall that $\bar\eps^I$ and $\bar\psi^I$ have raised index $I$ by default, and as that we rise that index by $\epsilon^{IJ}$ which is antisymmetric. We then get
\bea
{\mbox{fermionic bilinears}} &=& - 2 i r^2 \{\bar\psi \gamma_n \eps\bar\eps \gamma_m \chi\}\cr
&& + C_1 \{\bar\psi \eps\bar\eps\gamma_{mn}\chi\}\cr
&& - C_1 \{\bar\chi \eps\bar\eps\gamma_{mn}\psi\}
\eea
Then we expand 
\bea
\eps \bar\eps &=& c + c^p \gamma_p + c^{pq} \gamma_{pq}
\eea
where the first two coefficients are related to $s$ and $s^p$ by some factor. Here we just need the relation
\bea
c^p &=& \frac{s^p}{8}
\eea
We get
\bea
{\mbox{fermionic bilinears}} &=& - 2 i r^2 \{\bar\psi \gamma_n (c + c^p \gamma_p + c^{pq} \gamma_{pq}) \gamma_m \chi\}\cr
&& + C_1 \{\bar\psi  (c + c^p \gamma_p + c^{pq} \gamma_{pq}) \gamma_{mn}\chi\}\cr
&& - C_1 \{\bar\chi  (c + c^p \gamma_p + c^{pq} \gamma_{pq}) \gamma_{mn}\psi\}
\eea
We now need to apply flipping rules to the last line. The curly bracket is a reminder that we have antisymmetry due to Lie algebra commutator. Also the fermionic fields are anticommuting, there is a hidden index $I$ that is contracted by antisymmetric $c$ and $c^p$ and symmetric $c^{pq}$. Taking everything into account, we find the flipping rule
\bea
\{\bar\chi  (c + c^p \gamma_p + c^{pq} \gamma_{pq}) \gamma_{mn}\psi\} &=& 
\{\bar\psi \gamma_{mn} (- c - c^p \gamma_p - c^{pq} \gamma_{pq}) \chi\} 
\eea
and so we get
\bea
{\mbox{fermionic bilinears}} &=& - i r^2 \{\bar\psi (-2c \gamma_{mn} + c^p \{\gamma_p,\gamma_{mn}\} - c^{pq} \{\gamma_{pq},\gamma_{mn}\}) \chi\}\cr
&& + C_1 \{\bar\psi  \(2c\gamma_{mn} + c^p \{\gamma_p,\gamma_{mn}\} + c^{pq} \{\gamma_{pq},\gamma_{mn}\}\)\chi\}
\eea
By using that 
\bea
C_1 &=& - i r^2
\eea
we get
\bea
{\mbox{fermionic bilinears}} &=& - \frac{i r^2 s^p}{2} \{\bar\psi \gamma_{mnp} \chi\}
\eea
The selfdual Bianchi identity therefore becomes
\bea
3 D_{[r} \F_{mn]} + \frac{i n}{2 r} \E_{mnr}{}^{pq} \F_{pq} - \frac{ir}{2} \E_{mnr}{}^{pq} \([\sigma,f_{pq}]-[\phi,\F_{pq}]\) + \frac{r^2}{2} \{\bar\psi\gamma_{mnr}\chi\} &=& 0
\eea
This can be viewed as the standard Bianchi identity 
\bea
3D_{[r} \F_{mn]} - in H_{mnr} &=& 0
\eea
but with a modified selfduality condition
\bea
H_{mnr} &=& - \frac{1}{2 r} \E_{mnr}{}^{pq} \(\F_{pq} - \frac{r^2}{n} \([\sigma,f_{pq}]-[\phi,\F_{pq}]\) - \frac{i r^3}{2n} \{\bar\psi\gamma_{pq}\chi\}\)
\eea
where terms proportional to $1/n$ are nonlocal.

\subsection{Closure on $\chi$}
We carry out the closure computation on $\chi$ for each chiral component separately. We  decompose the gamma matrices into horizontal and vertical components, 
\bea
\gamma^m &=& \gamma'^m + \frac{s^m}{s} \gamma\cr
\gamma &=& \frac{s_m}{s} \gamma^m
\eea
and we put a prime on anything that is traceless (has vanishing contraction with $s^m$). The fields are decomposed into horizontal and vertical components as well,
\bea
\F_{mn} &=& \F'_{mn} + 2 \F'_{[m} \frac{s_{n]}}{s}\cr
\F'_m &=& \F_{mn} \frac{s^n}{s}
\eea
In the Appendix we show that
\bea
\gamma \eps &=& -\eps
\eea
Let us decompose the fermion field decomposes into its two chiral components
\bea
\chi &=& \chi^+ + \chi^-\cr
\gamma \chi^{\pm} &=& \pm \chi^{\pm}
\eea
These then will have the supersymmetry variations
\bea
\delta \chi^- &=& \frac{1}{2r} \gamma'^{mn} \eps \F'_{mn} - \frac{s^m}{s} \eps D_m \sigma + \frac{i n}{r} \eps \sigma + \frac{1}{r} C \eps [\sigma^{(0)},\sigma]\cr
\delta \chi^+ &=& - \frac{1}{r} \gamma'^m \eps \F'_m + \gamma'^m \eps D'_m \sigma
\eea
We also make a new ansatz for the variation of the field strength,
\bea
\delta \F'_{mn} &=& - 2 i r \bar\eps^I \gamma'_m D'_n \chi^+ - n \bar\eps \gamma'_{mn} \chi^- \cr
&& + i C_1 [\phi,\bar\eps \gamma'_{mn} \chi^-] - i C_1 [\sigma,\bar\eps \gamma'_{mn} \psi^-]\cr
\delta \F'_m &=& - i r \frac{s^n}{s} \bar\eps \gamma'_m D_n \chi^+ - n \bar\eps \gamma'_m \chi^+ - i r \bar\eps D'_m \chi^- \cr
&& + C' [\phi,\bar\eps \gamma'_m \chi^+] - C' [\sigma,\bar\eps^I \gamma'_m \psi^+]
\eea
that reduces to the above ansatz when we take $C' = r^2$, but it can be nice to see how this happens by closing supersymmetry on the fermions, so we keep $C'$ general for the moment.

\subsubsection{Closure on $\chi^+$}
\bea
\delta^2 \chi^+ &=& - i s^m D_m \chi^+ - \frac{ns}{r} \chi^+\cr
&& + \(\frac{C'}{r} - r\) \gamma'^m \eps \bar\eps \gamma'_m [\sigma,\psi^{+}]\cr
&& - \frac{C'}{r} \gamma'^m \eps \bar\eps \gamma'_m [\phi,\chi^+]
\eea
For the second line to vanish we need to take
\bea
C' &=& r^2
\eea
We use the Fierz identity
\bea
\epsilon \bar\epsilon &=& - \frac{s}{8} (1-\gamma) + \frac{1}{8} \Theta^{mn} \gamma'_{mn}
\eea
Here $\Theta$ is horizontal and selfdual. We also use the identities
\bea
\gamma'^m \gamma'_{pq} \gamma'_m &=& 0\cr
\gamma'^m \gamma'_m &=& 4
\eea
and then we get
\bea
\delta^2 \chi^+ &=& - i s^m D_m \chi^+ - \frac{ns}{r} \chi^+ - i [\chi^+,\lambda]
\eea
where
\bea
\lambda &=& - i r s \phi
\eea

\subsubsection{Closure on $\chi^-$}
\bea
\delta^2 \chi^- &=& - i \(\gamma'^{mn} \eps \bar\eps \gamma'_m - \frac{s^n}{s} \eps \bar\eps \) D_n \chi\cr
&& - \frac{n}{r} \(\frac{1}{2}\gamma'^{mn} \eps \bar\eps \gamma'_{mn} - \eps \bar\eps\) \chi^-\cr
&& + \(\frac{i C_1}{2 r} \gamma'^{mn} \eps \bar\eps \gamma'_{mn} - \frac{i C}{r}\eps \bar\eps \) [\phi,\chi^-]\cr
&& + \(\frac{i C_1}{2r} \gamma'^{mn} \eps \bar\eps \gamma'_{mn} + \(r - \frac{i C}{r}\) \eps \bar\eps\) [\psi^{-},\sigma]
\eea
We now use the following identity
\bea
\gamma'^{mn} \gamma'_{pq} \gamma'_m &=& \gamma'_{pq} \gamma'^n
\eea
that implies
\bea
\gamma'^{mn} \eps_I \bar\eps^J \gamma'_m &=& \frac{3 s}{4} \gamma'^n (P_+)_I{}^J + \frac{1}{8}\Theta^{pq}{}_I{}^J \gamma'_{pq} \gamma'^n\cr
\gamma'^{mn} \eps_I \bar\eps^J \gamma'_{mn} &=& 3 s (P_-)_I{}^J + \frac{1}{2} \Theta^{pq}{}_I{}^J \gamma'_{pq}
\eea
One can see that these are mutually consistent by contracting the first relation by $\gamma'^n$ from the right.
Anything involving $P_+$ goes into the equation of motion. We get
\bea
\delta^2 \chi^- &=& - i s^n D_n \chi^- - \frac{ns}{r} \chi^-\cr
&& - \frac{3i s}{4} \(\gamma'^n D'_n \chi^+ - \frac{s^n}{s} D_n \chi^- - \frac{i n}{r} \chi^-\)\cr
&& - \frac{is}{r} \(\frac{3C_1}{2} + \frac{C}{4}\) [\chi^-,\phi]\cr
&& + \frac{is}{r} \(\frac{3C_1}{2} + \frac{C}{4} + \frac{i r^2}{4}\) [\psi^-,\sigma]\cr
&& - \frac{i}{8} \Theta^{pq} \gamma_{pq} \(\gamma^n D_n \chi - \frac{s^n}{s} D_n \chi - \frac{in}{r} \chi\)\cr
&& - \frac{i}{8r} \Theta^{pq} \gamma_{pq} \(2 C_1 - C\) [\chi^-,\phi]\cr
&& + \frac{i}{8r} \Theta^{pq} \gamma_{pq} \(2 C_1 - C - i r^2\) [\psi_J^{-},\sigma]
\eea
Here, in the third and fourth lines, we put $C = C_1$ to get 
\bea
- \frac{7 i s C_1}{4 r} [\chi^-,\phi]\cr
+ \frac{7 i s C_1 - s r^2}{4 r} [\psi^-,\sigma]
\eea
In the third line, we then decompose
\bea
7 &=& 4 + 3
\eea
Then $4$ goes into the gauge transformation and $3$ goes into the equation of motion. Thus we get
\bea
\delta_{gauge} \chi^- &=& - i [\chi^-,\lambda]
\eea
with 
\bea
\lambda &=& \frac{s C_1}{r} \phi
\eea
and we get the equation of motion
\bea
\gamma'^n D'_n \chi^+ - \frac{s^n}{s} D_n \chi^- - \frac{i n}{r} \chi^- + \frac{C_1}{r} [\chi^-,\phi] + \frac{7 C_1 + i r^2}{3 r}[\psi^{-}_I,\sigma] &=& 0
\eea
The contributions that are proportional to $\Theta^{pq}$ are given by
\bea
- \frac{i}{8} \Theta^{pq} \gamma_{pq} \(\gamma^n D_n \chi - \frac{s^n}{s} D_n \chi - \frac{in}{r} \chi + \frac{C_1}{r} [\chi^-,\phi] + \frac{C_1 - i r^2}{r} [\psi,\sigma]\)
\eea
If we plug in the values
\bea
C = C_1 = - i r^2
\eea
the two fermionic equations of motion agree, since then
\bea
\frac{7 C_1 + i r^2}{3} &=& C_1 - i r^2
\eea

\section{The result} 
The following 5d supersymmetry variations, which consist of a SYM part (zero modes part),
\bea
\delta a_m &=& i r \bar\eps\gamma_m \psi\cr
\delta \phi &=& - i \bar\eps \psi\cr
\delta \psi &=& \frac{1}{2r} \gamma^{mn} \eps f_{mn} + \gamma^m \eps D_m \sigma
\eea
and a KK-mode part (nonzero mode part),
\ben
\delta \F_{mn} &=& - 2 i r \bar\eps\gamma_m D_n \chi - n \bar\eps \gamma_{mn} \chi + r^2 [\phi,\bar\eps\gamma_{mn}\chi] - r^2 [\sigma,\bar\eps\gamma_{mn}\psi]\cr
\delta \sigma &=& - i \bar\eps \chi\cr
\delta \chi &=& \frac{1}{2r} \gamma^{mn} \eps \F_{mn} + \gamma^m \eps D_m \sigma + \frac{i n}{r} \eps \sigma - i r \eps [\phi,\sigma]\label{gf}
\een
close on-shell on the fermionic equation of motion
\bea
\gamma^n D_n \chi - \frac{in}{r} \chi - i r [\chi,\phi] - 2 i r [\psi,\sigma] &=& 0
\eea
the Bianchi identity
\bea
3D_{[r} \F_{mn]} - in H_{mnr} &=& 0
\eea
and a modified selfduality condition
\ben
H_{mnr} &=& - \frac{1}{2 r} \E_{mnr}{}^{pq} \(\F_{pq} - \frac{r^2}{n} \([\sigma,f_{pq}]-[\phi,\F_{pq}]\) - \frac{i r^3}{2n} \{\bar\psi\gamma_{pq}\chi\}\)\label{sd}
\een

\section{The non-Abelian gerbe}
The gauge symmetry in 5d has an interesting non-Abelian gerbe structure \cite{Ho:2011ni}, \cite{Ho:2014eoa}, \cite{Huang:2018hho}. In the simplest set-up where all the fields transform in the adjoint representation, the gauge symmetry variations are
\bea
\delta \sigma &=& - i [\sigma,\lambda]\cr
\delta a_m &=& D_m \lambda
\eea
for the 5d SYM fields, and
\bea
\delta A_m &=& D_m \Lambda_0 - i n \Lambda_m - i [A_m,\lambda]\cr
\delta B_{mn} &=& 2 D_{[m} \Lambda_{n]} - i [B_{mn},\lambda] - i c [f_{mn},\Lambda_0]
\eea
for the 5d KK modes. Here the coefficient $c$ (that may depend on the mode number $n$) is not fixed by demanding closure of these gauge transformations alone. Instead this coefficient will be fixed below in a different way. The closure relation for these gauge variations reads
\bea
[\delta_{\Lambda'},\delta_{\Lambda}] &=& \delta_{\Lambda''}
\eea
where the new gauge parameters are given by 
\bea
\Lambda_m'' &=& -i\([\lambda',\Lambda_m] - [\lambda,\Lambda_m']\)\cr
\Lambda_0'' &=& -i\([\lambda',\Lambda_0] - [\lambda,\Lambda_0']\)\cr
\lambda' &=& -i[\lambda',\lambda]
\eea
These closure relations hold for any value of $c$. The YM field strength 
\bea
f_{mn} &=& \partial_m a_n - \partial_n a_m - i [a_m,a_n]\cr
\eea
transforms homogeneously
\bea
\delta f_{mn} &=& - i [f_{mn},\lambda]
\eea 
For the KK modes, we shall define the field strengths as
\ben
H_{mnp} &=& 3 D_m B_{np} - \frac{3}{n} [f_{mn},A_p]\cr
\F_{mn} &=& i n B_{mn} + 2 D_m A_n\label{fieldstrength}
\een
These transform homogeneously
\bea
\delta H_{mnp} &=& - i [H_{mnp},\lambda]\cr
\delta \F_{mn} &=& - i [\F_{mn},\lambda]
\eea
only for one particular value 
\bea
c &=& - \frac{1}{in}
\eea
and so this is what ultimately determines the value of this parameter in the gauge variations. One may now check that the Bianchi identity 
\bea
3D_{[r} \F_{mn]} - in H_{mnr} &=& 0
\eea
is satisfied. 

By using the modified selfduality condition (\ref{sd}), one can show that the supersymmetry variations \cite{Ho:2014eoa}
\bea
\delta \sigma &=& - i \bar\eps \chi\cr
\delta A_m &=& i r \bar\eps\gamma_m\chi\cr
\delta B_{mn} &=& i \bar\eps\gamma_{mn} \chi - \frac{i r^2}{n} \([\sigma,\bar\eps\gamma_{mn}\psi]-[\phi,\bar\eps\gamma_{mn}\chi]\) - \frac{2 i r}{n} [A_n,\bar\eps\gamma_m\psi]\cr
\delta \chi &=& \frac{1}{2r} \gamma^{mn} \eps \F_{mn} + \gamma^m \eps D_m \sigma + \frac{i n}{r} \eps \sigma - i r \eps [\phi,\sigma]
\eea
close on-shell up to a gerbe gauge transformation for each field $\Phi = (\sigma,A_m,B_{mn},\chi)$,
\bea
\delta^2 \Phi &=& - i s^m \partial_m \Phi - \frac{sn}{r} \Phi + \delta_{gauge} \Phi
\eea
with the gerbe gauge transformation parameters
\bea
\lambda &=& i s^m \(a_m - r \frac{s_m}{s}\phi\)\cr
\Lambda_m &=& \frac{is}{r} \(A_m - r \frac{s_m}{s} \sigma\) - i B_{mn} s^n + \frac{irs}{n} \[A_m - r \frac{s_m}{s} \sigma,\phi\]\cr
\Lambda_0 &=& i s^m \(A_m - r \frac{s_m}{s} \sigma\)
\eea
and that the supersymmetry variations (\ref{gf}) is a consequence of the above. What gets clarified when we express the supersymmetry variations in terms of the gauge potentials, is the underlying gerbe gauge structure and the fact that the modified selfdual Bianchi identity really shall be viewed as an ordinary Bianchi identity with the modified selfduality condition (\ref{sd}). This is in particular needed in order to obtain closure on $B_{mn}$.

\section{Discussion}
One may obtain off-shell supersymmetry variations, include 6d hypermultiplets, and try to enhance to $(2,0)$ superconformal symmetry. This is work in progress. 

Finally we should probably discuss the two proposals, \cite{Douglas:2010iu}, \cite{Lambert:2010iw} versus \cite{Bonetti:2012st}. Can both proposals be valid at the same time, or does one of them have to be wrong? For the proposal in \cite{Douglas:2010iu}, \cite{Lambert:2010iw}, few checks have been concerning the non-BPS sector of the theory. Maybe the conjecture is valid for the BPS sector, but as we go beyond the BPS sector maybe we will need another description for the non-Abelian tensor multiplet?

\subsection{Acknowledgments}
This work was supported in part by NRF Grant 2017R1A2B4003095. I would like to thank Kuo-Wei Huang for raising a critical question regarding the discussion section of the first version of this paper that made me realize I had made a mistake.

\appendix
\section{Spinor conventions}
The 5d spinors have been studied in for example \cite{Hosomichi:2012ek}, \cite{Kallen:2012va}. We have four-component spinors in 5d, whose spinor indices we usually suppress. But in this appendix, we will display all the indices. The spinor indices are denoted $\alpha,\beta,..$. We use the NW-SE convention to rise and lower indices,
\bea
v^{\alpha} &=& C^{\alpha\beta} v_{\beta}\cr
v_{\alpha} &=& v^{\beta} C_{\beta\alpha}
\eea
where $C_{\alpha\beta}$ is the charge conjugation matrix. It is antisymmetric. We define
\bea
C^{\alpha\beta} &=& C^{\alpha\alpha'} C^{\beta\beta'} C_{\alpha'\beta'}
\eea
Consistency requires that
\bea
C^{\alpha\beta} C_{\beta\gamma} &=& - \delta^{\alpha}_{\beta}
\eea
There are two supercharges, labeled by an index $I = 1,2$. These two supercharges form a doublet of the $SU(2)$ R symmetry. We use the same type of NW-SE convention for the index $I$,
\bea
v^I &=& \epsilon^{IJ} v_J\cr
v_I &=& v^J \epsilon_{JI}
\eea
and
\bea
\epsilon^{IJ} \epsilon_{JK} &=& - \delta^I_K
\eea
where $\epsilon_{IJ}$ is the antisymmetric tensor of $SU(2)$. Throughout this paper we use commuting supersymmetry parameters $\epsilon^{\alpha}_I$. We have the following Fierz identity
\bea
\eps^{\alpha}_I \eps^{\beta}_J &=& \frac{1}{8} \epsilon_{IJ} \(C^{\alpha\beta}s + \gamma^{\alpha\beta}_m s^m\) + \frac{1}{8} \gamma^{\alpha\beta}_{mn} \Theta_{IJ}^{mn}
\eea
Using the NW-SE convention, and 
\bea
C_{\alpha\beta} &=& - C_{\beta\alpha}\cr
\gamma^m_{\alpha\beta} &=& - \gamma^m_{\beta\alpha}\cr
\gamma^{mn}_{\alpha\beta} &=& \gamma^{mn}_{\beta\alpha}
\eea
we then get
\bea
s &=& \epsilon^{IJ} \eps_I^{\alpha} C_{\alpha\beta} \eps^{\beta}_J\cr
s^m &=& \epsilon^{IJ} \eps_I^{\alpha} \gamma^m_{\alpha\beta} \eps_J^{\beta}
\eea
and
\bea
\Theta^{mn}_{IJ} &=& \eps^{\alpha}_I \gamma^{pq}_{\alpha\beta} \eps^{\beta}_J
\eea
Upon contractions, we get
\bea
\epsilon^{IJ} \eps_I^{\alpha} \eps_J^{\beta} &=& \frac{s}{2} \P^{\alpha\beta}\cr
\eps_I^{\alpha} C_{\alpha\beta} \eps_J^{\beta} &=& \frac{s}{2} \eps_{IJ}
\eea
where
\bea
\P^{\alpha\beta} &=& \frac{1}{2} \(C^{\alpha\beta} + \gamma_m^{\alpha\beta} \frac{s^m}{s} \)
\eea
Now here is a subtlety. When we lower $\beta$ with $C_{\beta\gamma}$, we get
\bea
\P^{\alpha}{}_{\beta} &=& - P^{\alpha}{}_{\beta}
\eea
where we define
\bea
P^{\alpha}{}_{\beta} &=& \frac{1}{2} \(\delta^{\alpha}_{\beta} - \gamma^{\alpha}{}_{\gamma}\)\cr
\gamma^{\alpha}{}_{\beta} &=& (\gamma_m)^\alpha{}_\beta \frac{s^m}{s}
\eea
Using this, we get
\bea
P^{\alpha}{}_{\beta} \eps^{\beta}_I &=& \eps^{\alpha}_I
\eea
or 
\ben
\gamma^{\alpha}{}_{\beta} \eps^{\beta}_I &=& - \eps^{\alpha}_I\label{Weylminus}
\een
Let us check this is really consistent with our definitions. We have
\bea
s^m s_m &=& \epsilon^{IJ} \eps_I^{\alpha} \gamma^m_{\alpha\beta} s_m \eps_J^{\beta}
\eea
In order for the right-hand side to be equal to $s^2$, we must have
\bea
\gamma^m_{\alpha\beta} s_m \eps_J^{\beta} &=& s C_{\alpha\beta} \eps^{\beta}_J
\eea
Now rising $\alpha$ and using $C^{\alpha}{}_{\beta} = - \delta^{\alpha}_{\beta}$ we get the Weyl projection (\ref{Weylminus}).

We have the Fierz identity
\bea
(\gamma^m)_{\alpha\beta} (\gamma_m)_{\gamma\delta} &=& 2 C_{\delta\alpha} C_{\beta\gamma} - 2 C_{\delta\beta} C_{\alpha\gamma} - C_{\alpha\beta} C_{\gamma\delta}
\eea
Using this, we get
\bea
s^m s_m &=& s^2
\eea

We have 
\bea
\gamma_m &=& \gamma'_m + \gamma \frac{s_m}{s}\cr
\gamma_{mn} &=& \gamma'_{mn} + 2 \gamma'_m \gamma \frac{s_n}{s}
\eea
subject to the Clifford algebra
\bea
\{\gamma,\gamma'_m\} &=& 0\cr
\{\gamma'_m,\gamma'_n\} &=& 2 \(G_{mn} - \frac{s_m s_n}{s^2}\)
\eea
Using the identities
\bea
\gamma_{mnpqr} \gamma^{pq} &=& - 2 \gamma_{mnr}\cr
\gamma_{mnpqr} &=& \eps_{mnpqr}
\eea
we get upon contraction with $s^r$ the new identity
\bea
\gamma'_{mn} \gamma s &=& - \frac{1}{2} \eps_{mnpqr} \gamma'^{pq} s^r
\eea
By noting that 
\bea
\gamma_{mn} \frac{s^n}{s} &=& \gamma'_m \gamma
\eea

When we suppress indices, we use the following conventions, $
\eps = \eps_I^{\alpha}$ and $\bar\eps = \bar\eps^I_{\alpha} = \eps^T C = \epsilon^{IJ} \eps_J^{\beta} C_{\beta\alpha}$. For instance, in this notation, we have
\bea
\gamma \eps = - \eps
\eea
and by taking the transpose, we get
\bea
\eps^T C \gamma = - \eps^T C
\eea

Since we use the NW-SE rule, we have 
\bea
\gamma^m_{\alpha\beta} = (\gamma^m)^{\gamma}{}_{\beta} C_{\gamma\alpha} = - C_{\alpha\gamma} (\gamma^m)^{\gamma}{}_{\beta}
\eea
with an extra minus sign. This in turn means that 
\bea
s^m &=& \bar\eps^I \gamma^m \eps_I\cr
s &=& - \bar\eps^I \eps_I
\eea
This extra minus sign could have been avoided if we had chosen to instead using the SW-NE rule for the $\alpha$ index. In this paper, we will stick to the NW-SE rule for both the $\alpha$ and $I$ indices.

\section{The circle bundle}
Here we collect some circle bundle equations and relate 5d with 6d language following \cite{Linander:2011jy} which was concerned with the SYM part, but here we also consider  the KK modes. 

The 11d gamma matrices are denoted $\Gamma_M$ for $M=0,1,2,3,4,5$ and $\Gamma_A$ for $A=1,2,3,4,5$. The 6d tensor multiplet has the following supersymmetry variations
\bea
\delta \sigma &=& - i \bar\epsilon \chi\cr
\delta B_{MN} &=& i \bar\epsilon \Gamma_{MN} \chi\cr
\delta \chi &=& \frac{1}{12} \Gamma^{MNP} \epsilon H_{MNP} + \Gamma^M \epsilon \partial_M \sigma + 4 \eta \sigma
\eea
where the supersymmetry parameter satisfies the conformal Killing spinor equation
\bea
D_M \epsilon &=& \Gamma_M \eta
\eea
The 6d supersymmetry parameter and the spinor field are subject to the Weyl projections
\bea
\Gamma \eps &=& - \eps\cr
\Gamma \chi &=& \chi
\eea
where we define the 6d chirality matrix as
\bea
\Gamma &=& \Gamma^{\h 0 \h 1 \h 2 \h 3 \h 4 \h 5}
\eea
We define $H_{MNP} = 3\partial_{[M} B_{NP]}$. We find the following closure relations
\bea
\delta^2 \sigma &=& - i \L_S \sigma - \frac{i}{3} (D^M S_M) \sigma\cr
\delta^2 H_{MNP} &=& - i \L_S H_{MNP}\cr
\delta^2 \psi &=& - i \L_S \psi - \frac{5 i}{12} (D^M S_M) \psi\cr
&& + \frac{i}{4} S^M \Gamma_M \Gamma^N D_N \psi
\eea
where $\L_S$ denotes the Lie derivative along $S^M = \bar\epsilon \Gamma^M \epsilon$. Upon expanding these equations in flat space, we get closure relations that realize the  superconformal algebra.

We now consider the 6d tensor multiplet on a circle bundle with the metric 
\bea
ds^2 &=& - r^2 (dt + \kappa_m dx^m)^2 + G_{mn} dx^m dx^n
\eea
The vielbein is
\bea
e^{\h M}{}_M &=& \(\begin{array}{cc}
r & r \kappa_m\\
0 & E^{\h m}{}_m
\end{array}\)
\eea
with the inverse
\bea
e^{M}{}_{\h M} &=& \(\begin{array}{cc}
\frac{1}{r} & - \kappa_{\h m}\\
0 & E^m{}_{\h m}
\end{array}\)
\eea
From these we get the 6d metric components as
\bea
g_{mn} &=& G_{mn} - r^2 \kappa_m \kappa_n\cr
g_{m0} &=& - r^2 \kappa_m
\eea
and 
\bea
g^{mn} &=& G^{mn}\cr
g^{m0} &=& - \kappa^m\cr
g^{00} &=& - \frac{1}{r^2} + \kappa_m \kappa^m
\eea
where we define $\kappa^m = G^{mn} \kappa_n$ and $G_{mn}$ denote the 5d metric components.

We define the 6d and 5d spin connections from 
\bea
d e^{\h M} + \omega^{\h M \h N} \wedge e^{\h N} &=& 0\cr
d E^{\h m} + \Omega^{\h m \h n} \wedge E^{\h n} &=& 0
\eea
where
\bea
e^{\h 0} &=& r \(dt + \kappa\)\cr
e^{\h m} &=& E^{\h m}
\eea
We find
\bea
\omega^{\h m \h n}_{\h p} &=& \Omega^{\h m \h n}_{\h p}\cr
\omega^{\h m \h 0}_{\h 0} &=& \frac{1}{r} D_{\h m} r\cr
\omega^{\h m \h 0}_{\h n} &=& - \frac{r}{2} w_{\h m \h n}\cr
 \omega^{\h m \h n}_{\h 0} &=& \omega^{\h m \h 0}_{\h n}
\eea
Alternatively
\bea
\omega^{\h m \h n}_p &=& \Omega^{\h m \h n}_p - \frac{r^2}{2} \kappa_p w^{\h m \h n}\cr
\omega^{\h m \h n}_0 &=& - \frac{r^2}{2} w^{\h m \h n}\cr
\omega^{\h m \h 0}_0 &=& D^{\h m} r\cr
\omega^{\h m \h 0}_p &=& - \frac{r}{2} w^{\h m}{}_p + \kappa_p D^{\h m} r
\eea
Hence the covariant derivatives are
\bea
D_M \chi &=& \partial_M \chi + \frac{1}{4} \omega_M^{\h M \h N} \Gamma^{\h M \h N} \chi
\eea
We get
\bea
D_0 \chi &=& \partial_0 \chi - \frac{r^2}{8} w_{mn} \Gamma^{mn} \chi + \frac{1}{2} (D_m r) \Gamma^{m\h 0} \chi\cr
D_m \chi &=& \t\D_m \chi - \frac{r^2}{8} \kappa_m w_{pq} \Gamma^{pq} \chi + \frac{1}{2} \kappa_m D_n r \Gamma^{n \h 0} \chi + \frac{r}{4} w_{mn} \Gamma^{n\h 0} \chi\cr
&=& \t D_m \chi + \kappa_m \(D_0 - \partial_0\) \chi + \frac{r}{4} w_{mn} \Gamma^{n\h 0} \chi
\eea
where we define
\bea
\t \D_m \chi &=& \t D_m \chi - \kappa_m \partial_0 \chi
\eea
and we put tilde on 5d quantities, so for example
\bea
\t D_m = \partial_m + \frac{1}{4}\Omega_m^{\h m \h n} \Gamma^{\h m \h n}
\eea
On the other hand, the Lie derivative is given by
\bea
\L_V \chi &=& V^M D_M \chi + \frac{1}{8} W_{MN} \Gamma^{MN} \chi,\cr
W_{MN} &=& \partial_M V_N - \partial_N V_M
\eea
Then 
\bea
\L_V \chi &=& D_0 \chi + \frac{r^2}{8} w_{mn} \Gamma^{mn} \chi - \frac{1}{2} (D_m r) \Gamma^{m\h 0} \chi
\eea
so that when we expand out the spin connection, we get
\bea
\L_V \chi &=& \partial_0 \chi
\eea
We get
\bea
\bar\chi \Gamma^M D_M \chi &=& \frac{1}{r} \bar\chi \Gamma^{\h 0} \partial_0 \chi + \bar\chi \t\Gamma^m \t \D_m \chi + \frac{r}{8} w_{mn} \bar\chi \t\Gamma^{mn} \Gamma^{\h 0} \chi
\eea

We define 6d and 5d gamma matrices as 
\bea
\Gamma_M &=& \Gamma_{\h M} e^{\h M}{}_M\cr
\t \Gamma_m &=& \Gamma_{\h m} E^{\h m}{}_m
\eea
These are related as 
\bea
\Gamma_m &=& \t \Gamma_m + r \kappa_m \Gamma_{\h 0}\cr
\Gamma_0 &=& r \Gamma_{\h 0}
\eea
and
\bea
\Gamma^m &=& \t \Gamma^m\cr
\Gamma^0 &=& \frac{1}{r} \Gamma^{\h 0} - \kappa_m \t \Gamma^m
\eea

We define 
\bea
\Gamma^{MNPQRS} &=& \eps^{MNPQRS} \Gamma
\eea
We now use the gamma matrix identity
\bea
\Gamma^{MNPQRS} \Gamma_{QRS} &=& - 6 \Gamma^{MNP}
\eea
together with the above definition, to get
\bea
\frac{1}{6} \eps^{MNP}{}_{QRS} \Gamma^{QRS} &=& \Gamma^{MNP} \Gamma
\eea
Now let us define
\bea
H_{MNP}^{\pm} &=& \frac{1}{2} \(H_{MNP} \pm \frac{1}{6}\eps_{MNP}{}^{QRS}H_{QRS}\)
\eea
These are subject to 
\bea
H_{MNP}^{\pm} &=& \pm \frac{1}{6} \eps_{MNP}{}^{QRS} H_{QRS}^{\pm}
\eea
We have
\bea
\Gamma^{MNP} \eps H_{MNP} &=& \Gamma^{MNP} \eps H_{MNP}^+
\eea
and thus the relevant piece of the field strenght is the selfdual part that belongs to the tensor multiplet. It is subject to the selfduality constraint
\bea
H_{MNP}^+ &=& \frac{1}{6} \eps_{MNP}{}^{QRS} H_{QRS}^+
\eea

Now we will reduce this selfduality constraint to 5d. We start by defining the 6d covariant epsilon tensors
\bea
\eps_{M_1\cdots M_6} &=& e^{\h M_1}{}_{M_1} \cdots e^{\h M_6}{}_{M_6} \epsilon_{\h M_1\cdots \h M_6}\cr
&=& \sqrt{-g} \epsilon_{\h M_1\cdots \h M_6}\cr
\eps^{M_1\cdots M_6} &=& e^{M_1}{}_{\h M_1} \cdots e^{M_6}{}_{\h M_6} \epsilon^{M_1\cdots M_6}\cr
&=& \frac{1}{\sqrt{-g}} \epsilon^{M_1\cdots M_6}
\eea
where 
\bea
\epsilon^{012345} &=& 1\cr
\epsilon_{012345} &=& -1
\eea
Likewise, we define the 5d covariant epsilon tensors as
\bea
\E_{m_1 \cdots m_5} &=& E^{\h m_1}{}_{m_1} \cdots E^{\h m_5}{}_{m_5} \epsilon_{\h m_1 \cdots \h m_5}\cr
&=& \sqrt{G} \epsilon_{\h m_1 \cdots \h m_5}\cr
\E^{m_1 \cdots m_5} &=& E^{m_1}{}_{\h m_1} \cdots E^{m_5}{}_{\h m_5} \epsilon^{\h m_1 \cdots \h m_5}\cr
&=& \frac{1}{\sqrt{G}} \epsilon^{m_1 \cdots m_5}
\eea
where 
\bea
\epsilon^{12345} &=& 1
\eea
By noting that 
\bea
\sqrt{-g} &=& r \sqrt{G}
\eea
we get the following relations
\bea
\eps_{0 m_1\cdots m_5} &=& - r \E_{m_1\cdots m_5}\cr
\eps^{0 m_1\cdots m_5} &=& - \frac{1}{r} \E^{m_1 \cdots m_5}
\eea
We have
\bea
H_{mn0} &=& \frac{1}{6} \eps_{mn0}{}^{qrs} H_{qrs} + \frac{1}{2} \epsilon_{mn0}{}^{qr0} H_{qr0}\cr
&=& - \frac{r}{6} \E_{mn}{}^{qrs} H_{qrs} + \frac{r}{2} \E_{mn}{}^{qrs} H_{qr0} \kappa_s\cr
&=& - \frac{r}{6} \E_{mn}{}^{qrs} \(H_{qrs} - 3 H_{qr0}\kappa_s\)
\eea
that we can invert,
\bea
H_{qrs} &=& - \frac{1}{2r} \E_{qrs}{}^{mn} H_{mn0} + 3 H_{[qr|0|} \kappa_{s]}
\eea

We have
\bea
\Gamma^{mn0} &=& \frac{1}{r} \t\Gamma^{mn} \Gamma^{\h 0} - \t\Gamma^{mn} \t\Gamma^p \kappa_p\cr
\Gamma^{mnp} &=& \t\Gamma^{mnp}
\eea
and also 
\bea
\Gamma^{mn} \Gamma^0 + \Gamma^0 \Gamma^{mn} + \Gamma^n \Gamma^0 \Gamma^m &=& \frac{3}{r} \t\Gamma^{mn} \Gamma^{\h 0} - 3\t\Gamma^{mnp} \kappa_p
\eea
which leads to
\bea
3\Gamma^{mn0} \eps H_{mn0} &=& \frac{3}{r} \t \Gamma^{mn} \Gamma^{\h 0} \eps H_{mn0} - 3 \t\Gamma^{mnp} \eps H_{mn0} \kappa_p\cr
\Gamma^{mnp} \eps H_{mnp} &=& - \frac{1}{2r} \t\Gamma^{mnp} \eps \E_{mnp}{}^{qr} H_{qr0} + 3 \t\Gamma^{mnp} \eps H_{mn0} \kappa_p
\eea
Adding these, we find a nice cancelation,
\bea
\frac{1}{6} \Gamma^{MNP} \eps H_{MNP} &=& \frac{1}{r} \t\Gamma^{mn} \Gamma^{\h 0} \eps H_{mn0}
\eea
We have
\bea
\Gamma^M \eps D_M \sigma &=& \t\Gamma^m \eps \(D_m \sigma - \kappa_m D_0 \sigma\) + \frac{1}{r} \Gamma^{\h 0} \eps D_0 \sigma
\eea
We have 
\bea
\Gamma_{mn} &=& \t\Gamma_{mn} + 2 r \kappa_{[m} \Gamma_{\h 0}\t \Gamma_{n]}\cr
\Gamma_{m0} &=& r \t\Gamma_m\Gamma_{\h 0} - r^2 \kappa_m
\eea

We are now ready to decompose the Abelian supersymmetry variations into one zero mode part. We define
\bea
a_m &=& B_{m0}^{(0)}
\eea
Then we get
\bea
\delta \sigma^{(0)} &=& - i \bar\eps \chi^{(0)}\cr
\delta a_m &=& i r \bar\eps\t\Gamma_{m}\Gamma_{\h 0} \chi^{(0)} - i r^2 \kappa_m \bar\eps \chi^{(0)}\cr
\delta \chi^{(0)} &=& \frac{1}{2r} \t\Gamma^{mn} \Gamma^{\h 0} \eps F_{mn} + \t\Gamma^m\eps D_m\sigma^{(0)} + 4 \eta \sigma^{(0)}
\eea
For the KK-modes, we get
\bea
\delta \sigma^{(n)} &=& - i \bar\eps \chi^{(n)}\cr
\delta B_{m0}^{(n)} &=& i r \bar\eps \t\Gamma_m \Gamma_{\h 0} \chi^{(n)} - i r^2 \kappa_m \bar\eps \chi^{(n)}\cr
\delta B_{mn}^{(n)} &=& i \bar\eps \t\Gamma_{mn} \chi^{(n)} + 2 i r \kappa_{[m} \bar\eps \Gamma_{\h 0}\t\Gamma_{n]} \chi^{(n)}\cr
\delta \chi^{(n)} &=& \frac{1}{2r} \t\Gamma^{mn} \Gamma^{\h 0}\eps H_{mn0}^{(n)} + \t\Gamma^m \eps \(D_m \sigma^{(n)} - \kappa_m D_0 \sigma^{(n)}\) + \frac{1}{r} \Gamma^{\h 0} \eps D_0 \sigma^{(n)} + 4 \eta \sigma^{(n)}
\eea
Now we use
\bea
\Gamma^{\h 0} &=& i \sigma^2 \otimes 1\otimes 1\cr
\Gamma^{\h m} &=& \sigma^1 \otimes \gamma^{\h m} \otimes 1
\eea
We have 
\bea
\Gamma &=& \sigma^3 \otimes 1 \otimes 1
\eea
We have $\Gamma \chi = \chi$, so we get
\bea
\Gamma^{\h 0} \chi &=& - \chi\cr
\t\Gamma^{\h m} \chi &=& \gamma^m \chi
\eea
We get
\bea
\delta \sigma^{(0)} &=& - i \bar\eps \chi^{(0)}\cr
\delta a_m &=& i r \bar\eps \gamma_{m} \chi^{(0)} - i r^2 \kappa_m \bar\eps \chi^{(0)}\cr
\delta \chi^{(0)} &=& \frac{1}{2r} \gamma^{mn} \eps f_{mn} + \gamma^m\eps D_m\sigma^{(0)} + 4 \eta \sigma^{(0)}
\eea
and 
\bea
\delta \sigma^{(n)} &=& - i \bar\eps \chi^{(n)}\cr
\delta B_{m0}^{(n)} &=& i r \bar\eps \gamma_m \chi^{(n)} - i r^2 \kappa_m \bar\eps \chi^{(n)}\cr
\delta B_{mn}^{(n)} &=& i \bar\eps \gamma_{mn} \chi^{(n)} - 2 i r \kappa_{[m} \bar\eps \gamma_{n]} \chi^{(n)}\cr
\delta \chi^{(n)} &=& \frac{1}{2r} \gamma^{mn} \eps H_{mn0}^{(n)} + \gamma^m \eps \(D_m \sigma^{(n)} - i n\kappa_m \sigma^{(n)}\) + \frac{in}{r} \eps \sigma^{(n)} + 4 \eta \sigma^{(n)}
\eea
By putting $\kappa_m$ to zero, these supersymmetry variations reduce to those that appear in the main text for Abelian gauge group.

\end{document}